\documentclass[10pt]{article}
\usepackage{graphicx}
\usepackage{amsmath}
\usepackage{amssymb}
\usepackage{caption2}
\begin{document}
\begin{center}
{\large {\bf \sc{  Lowest  vector tetraquark states: $Y(4260/4220)$ or $Z_c(4100)$  }}} \\[2mm]
Zhi-Gang  Wang \footnote{E-mail: zgwang@aliyun.com.  }     \\
 Department of Physics, North China Electric Power University, Baoding 071003, P. R. China
\end{center}

\begin{abstract}
In this article, we take the $Y(4260/4220)$ as the vector tetraquark state with $J^{PC}=1^{--}$, and construct  the $C\gamma_5\otimes\stackrel{\leftrightarrow}{\partial}_\mu\otimes \gamma_5C$ type diquark-antidiquark current to study its mass and pole residue with the QCD sum rules in details  by taking into account the vacuum condensates up to dimension 10 in a consistent way. The predicted mass $M_{Y}=4.24\pm0.10\,\rm{GeV}$ is in excellent agreement with experimental data and supports  assigning the $Y(4260/4220)$  to be the  $C\gamma_5\otimes\stackrel{\leftrightarrow}{\partial}_\mu\otimes \gamma_5C$  type vector tetraquark state, and disfavors assigning the $Z_c(4100)$  to be the  $C\gamma_5\otimes\stackrel{\leftrightarrow}{\partial}_\mu\otimes \gamma_5C$  type vector tetraquark state. It is the first time that the QCD sum rules have reproduced  the mass of the $Y(4260/4220)$ as a vector tetraquark state.
 \end{abstract}



\section{Introduction}

In 2005, the BaBar collaboration   observed the $Y(4260)$   in the $\pi^+\pi^- J/\psi$ mass spectrum in the initial-state radiation process  $e^+ e^- \to \gamma_{ISR} \pi^+\pi^- J/\psi$ \cite{BaBar4260-0506}. Then the $Y(4260)$ was confirmed by the   Belle and CLEO collaborations  \cite{Belle-0707,CLEO-0606}.
There have been several possible assignments for the $Y(4260)$ since its observation, such as the tetraquark state \cite{Maiani-4260,Maiani-II-type,Ali-Maiani-Y,Brodsky-PRL,ZhangHuang-PRD,ZhangHuang-JHEP,Nielsen-4260-4460,WangEPJC-1601}, hybrid states \cite{Hybrid-4260,Hybrid-4260-Lattice,BO-potential}, hadro-charmonium state \cite{Voloshin-Y4260},
 molecular state \cite{Zhao-PRL-Y4260-3900-2,WangCPC-Y4390},  kinematical effect \cite{Chen-He-Liu-4260,CC-Effects}, baryonium state \cite{Qiao-CF-4260}, etc.

In 2014, the BES collaboration observed a resonance in the $\omega\chi_{c0}$ cross section in the processes  $e^+e^-\to \omega\chi_{c0/c1/c2}$,   the measured mass and width  are $4230\pm 8\pm 6\, \rm{ MeV}$   and $ 38\pm 12\pm 2\,\rm{MeV}$, respectively \cite{BES-2014-4230}.
In 2016, the BES collaboration observed the $Y(4220)$ and $Y(4390)$ in the  process $e^+ e^- \to \pi^+\pi^- h_c$, the measured masses  and widths are $M_{Y(4220)}=4218.4\pm4.0\pm0.9\,\rm{MeV}$, $M_{Y(4390)}=4391.6\pm6.3\pm1.0\,\rm{MeV}$, $\Gamma_{Y(4220)}=66.0\pm9.0\pm0.4\,\rm{MeV}$ and $\Gamma_{Y(4390)}=139.5\pm16.1\pm0.6\,\rm{MeV}$, respectively \cite{BES-Y4390}.
Also in 2016, the BES collaboration  observed the $Y(4220)$ and $Y(4320)$ by precisely measuring the cross section of the  process $e^+ e^- \to  \pi^+\pi^- J/\psi$,
the measured masses  and widths are $M_{Y(4220)}=4222.0\pm3.1\pm 1.4\,  \rm{MeV}$, $M_{Y(4320)}=4320.0\pm 10.4 \pm 7.0\, \rm{MeV}$,   $\Gamma_{Y(4220)}=44.1 \pm 4.3\pm 2.0 \,\rm{MeV}$ and $\Gamma_{Y(4320)}=101.4^{+25.3}_{-19.7}\pm 10.2\,\rm{MeV}$, respectively \cite{BES-Y4220-Y4320}. The $Y(4260)$ and $Y(4220)$ may be the same particle,  while the $Y(4360)$ and $Y(4320)$ may be the same particle according to the analogous masses and widths.

In Ref.\cite{Maiani-4260}, L. Maiani et al assign the $Y(4260)$ to be the  diquark-antidiquark type tetraquark state with the angular momentum $L=1$  based on the effective  Hamiltonian with the spin-spin and spin-orbit  interactions.
In the type-II diquark-antidiquark model \cite{Maiani-II-type}, where the spin-spin interactions between the quarks and antiquarks are neglected, L. Maiani et al interpret the $Y(4008)$, $Y(4260)$, $Y(4290/4220)$  and $Y(4630)$ as the four ground  states with  $L=1$. By incorporating the dominant spin-spin, spin-orbit and tensor interactions, A. Ali et al observe that the preferred  assignments of the ground state tetraquark states with  $L=1$ are the $Y(4220)$, $Y(4330)$, $Y(4390)$, $Y(4660)$ rather than the  $Y(4008)$, $Y(4260)$, $Y(4360)$, $Y(4660)$ \cite{Ali-Maiani-Y}.  The QCD sum rules can reproduce the experimental values of the masses of the $Y(4360)$ and $Y(4660)$ in the  scenario of the tetraquark states \cite{ZhangHuang-PRD,ZhangHuang-JHEP,Nielsen-4260-4460,WangEPJC-1601,Wang-tetra-formula,ChenZhu,Nielsen4260-1209,WangY4360Y4660-1803}.

The diquarks $\varepsilon^{ijk}q^{T}_j C\Gamma q^{\prime}_k$ have  five  structures  in Dirac spinor space, where $C\Gamma=C\gamma_5$, $C$, $C\gamma_\mu \gamma_5$,  $C\gamma_\mu $ and $C\sigma_{\mu\nu}$ for the scalar, pseudoscalar, vector, axialvector  and  tensor diquarks, respectively, the $i$, $j$, $k$ are color indexes.
The attractive interactions of one-gluon exchange  favor  formation of
the diquarks in  color antitriplet, flavor
antitriplet and spin singlet \cite{One-gluon},
 while the favored configurations are the scalar ($C\gamma_5$) and axialvector ($C\gamma_\mu$) diquark states based on the QCD sum rules \cite{WangDiquark,WangLDiquark,Tang-Diquark,Dosch-Diquark-1989}. We can take the $C\gamma_5$ and $C\gamma_\mu$ diquark states as basic constituents to construct the
 scalar and axialvector tetraquark states \cite{WangHuangtao-2014-PRD,Wang-tetra-NPA}. In the non-relativistic quark models, we have to introduce additional P-waves explicitly to study the vector tetraquark states, while in the quantum field theory, we can also take other diquark states ($C$, $C\gamma_\mu \gamma_5$  and $C\sigma_{\mu\nu}$) as basic constituents without introducing the explicit P-waves to  study the vector tetraquark states \cite{ZhangHuang-PRD,ZhangHuang-JHEP,Nielsen-4260-4460,WangEPJC-1601,Wang-tetra-formula,ChenZhu,WangJPG,Azizi-V}.
However, up to now, the QCD sum rules cannot reproduce the experimental value of the mass of the $Y(4260/4220)$ in the  scenario of the tetraquark state \cite{ZhangHuang-PRD,ZhangHuang-JHEP,Nielsen-4260-4460,WangEPJC-1601,Wang-tetra-formula,ChenZhu,Nielsen4260-1209,WangY4360Y4660-1803}. We often obtain much larger mass than the $M_{Y(4260/4220)}$.

The net effects of the relative P-waves between the heavy (anti)quarks and light (anti)quarks in the heavy (anti)diquarks are embodied in the underlined  $\gamma_5$ in the  $C\gamma_5 \underline{\gamma_5} \otimes \gamma_\mu C$ type and
$C\gamma_5 \otimes \underline{\gamma_5}\gamma_\mu C$ type currents or in the underlined  $\gamma^\alpha$ in the  $C\gamma_\alpha \underline{\gamma^\alpha} \otimes \gamma_\mu C$ type  currents \cite{WangY4360Y4660-1803}. If we introduce the relative P-waves  between the heavy (anti)quarks and light (anti)quarks in the heavy (anti)diquarks explicitly,
we can obtain the $\stackrel{\leftrightarrow}{\partial}_\mu C\gamma_5\otimes \gamma_5C$ type, $C\gamma_5\otimes\stackrel{\leftrightarrow}{\partial}_\mu \gamma_5C$ type,  $\stackrel{\leftrightarrow}{\partial}_\mu C\gamma_\alpha \otimes \gamma^\alpha C$ type or $C\gamma_\alpha \otimes \stackrel{\leftrightarrow}{\partial}_\mu \gamma^\alpha C$ type vector currents, for example,
$\varepsilon^{ijk}u^{Tj}(x)\stackrel{\leftrightarrow}{\partial}_{\mu}C\gamma_5 c^k(x)\,\varepsilon^{imn}\bar{d}^m(x)\gamma_5 C \bar{c}^{Tn}(x)$,
$\varepsilon^{ijk}u^{Tj}(x)C\gamma_5 c^k(x)\, \varepsilon^{imn}  \bar{d}^m(x)\stackrel{\leftrightarrow}{\partial}_\mu\gamma_5 C \bar{c}^{Tn}(x) $, where $\stackrel{\leftrightarrow}{\partial}_\mu=\stackrel{\rightarrow}{\partial}_\mu-\stackrel{\leftarrow}{\partial}_\mu$.
On the other hand, we  can introduce the relative P-waves between diquark and antidiquark  explicitly and construct the $C\gamma_5\otimes\stackrel{\leftrightarrow}{\partial}_\mu\otimes \gamma_5C$ type and $C\gamma_\alpha \otimes \stackrel{\leftrightarrow}{\partial}_\mu \otimes\gamma^\alpha C$ type currents to interpolate the vector tetraquark states \cite{WangZG-CPL}, for example,
$\varepsilon^{ijk}u^{Tj}(x)C\gamma_5 c^k(x)\stackrel{\leftrightarrow}{\partial}_\mu \varepsilon^{imn}\bar{d}^m(x)\gamma_5 C \bar{c}^{Tn}(x)$.

 The masses of the $C\gamma_5 \underline{\gamma_5} \otimes \gamma_\mu C$ type, $C\gamma_5 \otimes \underline{\gamma_5}\gamma_\mu C$, $\gamma^\alpha$ type,  $C\gamma_\alpha \underline{\gamma^\alpha} \otimes \gamma_\mu C$ type, $\stackrel{\leftrightarrow}{\partial}_\mu C\gamma_5\otimes \gamma_5C$ type, $C\gamma_5\otimes\stackrel{\leftrightarrow}{\partial}_\mu \gamma_5C$ type,  $\stackrel{\leftrightarrow}{\partial}_\mu C\gamma_\alpha \otimes \gamma^\alpha C$ type and $C\gamma_\alpha \otimes \stackrel{\leftrightarrow}{\partial}_\mu \gamma^\alpha C$ type  vector tetraquark states  maybe  differ from the masses of the
$C\gamma_5\otimes\stackrel{\leftrightarrow}{\partial}_\mu\otimes \gamma_5C$ type and $C\gamma_\alpha \otimes \stackrel{\leftrightarrow}{\partial}_\mu
 \otimes\gamma^\alpha C$ type vector tetraquark states greatly. In Refs.\cite{ZhangHuang-PRD,ZhangHuang-JHEP}, Zhang and Huang construct the $C\gamma_5\otimes\partial_\mu\otimes \gamma_5C$ type and $C\gamma_\alpha \otimes \partial_\mu \otimes\gamma^\alpha C$ type vector interpolating  currents, which have no definite charge conjugation,    and study the vector tetraquark states with the QCD sum rules  by taking into account  the vacuum condensates up to dimension 6  in the operator product expansion, and obtain the masses $4.32\,\rm{GeV}$ and $4.69\,\rm{GeV}$ for the $Y(4360)$ and $Y(4660)$ respectively.

In this article, we take the $Y(4260/4220)$ as the vector tetraquark state with the $J^{PC}=1^{--}$, and construct  the $C\gamma_5\otimes\stackrel{\leftrightarrow}{\partial}_\mu\otimes \gamma_5C$ type current to study its mass and pole residue with the QCD sum rules in details  by taking into account the vacuum condensates up to dimension 10 in a consistent way in the operator product expansion, and use the energy scale formula $\mu=\sqrt{M^2_{X/Y/Z}-(2{\mathbb{M}}_c)^2}$ with the effective $c$-quark mass ${\mathbb{M}}_c$ to determine the optimal energy scale of the QCD spectral density \cite{Wang-tetra-formula,WangHuangtao-2014-PRD,Wang-tetra-NPA,WangHuang-molecule}.

Recently, the LHCb collaboration observed evidence for the $\eta_c \pi^-$ resonant state $Z_c(4100)$ with the significance of  more than three standard deviations in a Dalitz plot analysis of the  $B^0 \to \eta_c  K^+\pi^- $ decays, the measured mass and width are $M_{Z_c}=4096 \pm 20^{+18}_{-22}\,\rm{MeV}$ and $\Gamma_{Z_c}= 152 \pm 58^{+60}_{-35}\,\rm{MeV}$ respectively \cite{LHCb-Z4100}. The spin-parity assignments $J^P =0^+$   and $1^-$   are both consistent with the experimental data. It is interesting to see which is the lowest vector tetraquark state,  the $Y(4260/4220)$ or the $Z_c(4100)$ ?

The article is arranged as follows:  we derive the QCD sum rules for the mass and pole residue  of  the vector   tetraquark state $Y(4260/4220)$ in section 2; in section 3, we   present the numerical results and discussions; section 4 is reserved for our conclusion.

\section{QCD sum rules for  the  vector tetraquark state $Y(4260/4220)$}
In the following, we write down  the two-point correlation function $\Pi_{\mu\nu}(p)$  in the QCD sum rules,
\begin{eqnarray}
\Pi_{\mu\nu}(p)&=&i\int d^4x e^{ip \cdot x} \langle0|T\left\{J_\mu(x)J_\nu^{\dagger}(0)\right\}|0\rangle \, ,
\end{eqnarray}
where $J_\mu(x)=J_\mu^+(x)$, $J_\mu^0(x)$ and $J_\mu^-(x)$,
\begin{eqnarray}
J^+_\mu(x)&=&\frac{\varepsilon^{ijk}\varepsilon^{imn}}{\sqrt{2}}u^{Tj}(x)C\gamma_5 c^k(x)\stackrel{\leftrightarrow}{\partial}_\mu \bar{d}^m(x)\gamma_5 C \bar{c}^{Tn}(x) \, ,\nonumber \\
J^0_\mu(x)&=&\frac{\varepsilon^{ijk}\varepsilon^{imn}}{2}\Big[u^{Tj}(x)C\gamma_5 c^k(x)\stackrel{\leftrightarrow}{\partial}_\mu \bar{u}^m(x)\gamma_5 C \bar{c}^{Tn}(x)\nonumber\\
&&\pm \,d^{Tj}(x)C\gamma_5 c^k(x)\stackrel{\leftrightarrow}{\partial}_\mu \bar{d}^m(x)\gamma_5 C \bar{c}^{Tn}(x) \Big] \, , \nonumber\\
J^{-}_\mu(x)&=&\frac{\varepsilon^{ijk}\varepsilon^{imn}}{\sqrt{2}}d^{Tj}(x)C\gamma_5 c^k(x)\stackrel{\leftrightarrow}{\partial}_\mu \bar{u}^m(x)\gamma_5 C \bar{c}^{Tn}(x) \, ,
\end{eqnarray}
where the $i$, $j$, $k$, $m$, $n$ are color indexes.  Under charge conjugation transform $\widehat{C}$, the currents $J_\mu(x)$ have the property,
\begin{eqnarray}
\widehat{C}J_{\mu}(x)\widehat{C}^{-1}&=&- J_\mu(x) \, .
\end{eqnarray}
We take the isospin limit by assuming the $u$ and $d$ quarks have  degenerate masses, the $J_\mu(x)$ couple to the vector tetraquark states with degenerate masses. In this article, we take $J_\mu(x)=J^+_\mu(x)$.

At the hadronic side, we can insert  a complete set of intermediate hadronic states with
the same quantum numbers as the current operator $J_\mu(x)$ into the
correlation function $\Pi_{\mu\nu}(p)$  to obtain the hadronic representation
\cite{SVZ79,Reinders85}. After isolating the ground state
contribution of the vector tetraquark state  $Y(4260/4220)$,  we get the result,
\begin{eqnarray}
\Pi_{\mu\nu}(p)&=&\frac{\lambda_{Y}^2}{M_{Y}^2-p^2}\left(-g_{\mu\nu} +\frac{p_\mu p_\nu}{p^2}\right) +\cdots \, \, ,\nonumber\\
&=&\Pi(p^2)\left(-g_{\mu\nu} +\frac{p_\mu p_\nu}{p^2}\right) +\Pi_0(p^2)\frac{p_\mu p_\nu}{p^2} \, ,
\end{eqnarray}
where the pole residue  $\lambda_{Y}$ is  defined by $\langle 0|J_\mu(0)|Y(p)\rangle=\lambda_{Y} \,\varepsilon_\mu$,
the $\varepsilon_\mu$ is the polarization vector of the  vector tetraquark state $Y(4260/4220)$. The vector and scalar tetraquark states contribute to the components $\Pi(p^2)$ and $\Pi_0(p^2)$, respectively. In this article, we choose the tensor structure $-g_{\mu\nu} +\frac{p_\mu p_\nu}{p^2}$ for analysis, the scalar tetraquark states have no contaminations.

 Now we briefly outline  the operator product expansion for the correlation function $\Pi_{\mu\nu}(p)$  in perturbative QCD.  We contract the $u$, $d$ and $c$ quark fields in the correlation function $\Pi_{\mu\nu}(p)$ with Wick theorem, obtain the result:
\begin{eqnarray}
\Pi_{\mu\nu}(p)&=&-\frac{i\varepsilon^{ijk}\varepsilon^{imn}\varepsilon^{i^{\prime}j^{\prime}k^{\prime}}\varepsilon^{i^{\prime}m^{\prime}n^{\prime}}}{2}\int d^4x e^{ip \cdot x}   \nonumber\\
&&\left\{{\rm Tr}\left[\gamma_5 C^{kk^{\prime}}(x)\gamma_5 CS^{jj^{\prime}T}(x)C\right] \partial_\mu \partial_\nu{\rm Tr}\left[ \gamma_5 C^{n^{\prime}n}(-x)\gamma_5 C S^{m^{\prime}mT}(-x)C\right] \right. \nonumber\\
&&-\partial_\mu{\rm Tr}\left[ \gamma_5 C^{kk^{\prime}}(x)\gamma_5 CS^{jj^{\prime}T}(x)C\right] \partial_\nu{\rm Tr}\left[ \gamma_5 C^{n^{\prime}n}(-x)\gamma_5 C S^{m^{\prime}mT}(-x)C\right] \nonumber\\
&&-\partial_\nu{\rm Tr}\left[ \gamma_5 C^{kk^{\prime}}(x)\gamma_5 CS^{jj^{\prime}T}(x)C\right] \partial_\mu{\rm Tr}\left[ \gamma_5 C^{n^{\prime}n}(-x)\gamma_5 C S^{m^{\prime}mT}(-x)C\right] \nonumber\\
 &&\left.+\partial_\mu \partial_\nu{\rm Tr}\left[ \gamma_5 C^{kk^{\prime}}(x)\gamma_5 CS^{jj^{\prime}T}(x)C\right] {\rm Tr}\left[  \gamma_5C^{n^{\prime}n}(-x)\gamma_5 C S^{m^{\prime}mT}(-x)C\right] \right\}\, ,
\end{eqnarray}
where  the $S_{ij}(x)$ and $C_{ij}(x)$ are the full $u/d$ and $c$ quark propagators respectively,
 \begin{eqnarray}
S_{ij}(x)&=& \frac{i\delta_{ij}\!\not\!{x}}{ 2\pi^2x^4}
-\frac{\delta_{ij}\langle\bar{q}q\rangle}{12} -\frac{\delta_{ij}x^2\langle \bar{s}g_s\sigma Gs\rangle}{192}-\frac{ig_s G^{a}_{\alpha\beta}t^a_{ij}(\!\not\!{x}
\sigma^{\alpha\beta}+\sigma^{\alpha\beta} \!\not\!{x})}{32\pi^2x^2}\nonumber\\
&&   -\frac{\delta_{ij}x^4\langle \bar{q}q \rangle\langle g_s^2 GG\rangle}{27648}-\frac{1}{8}\langle\bar{q}_j\sigma^{\mu\nu}q_i \rangle \sigma_{\mu\nu}   +\cdots \, ,
\end{eqnarray}
\begin{eqnarray}
C_{ij}(x)&=&\frac{i}{(2\pi)^4}\int d^4k e^{-ik \cdot x} \left\{
\frac{\delta_{ij}}{\!\not\!{k}-m_c}
-\frac{g_sG^n_{\alpha\beta}t^n_{ij}}{4}\frac{\sigma^{\alpha\beta}(\!\not\!{k}+m_c)+(\!\not\!{k}+m_c)
\sigma^{\alpha\beta}}{(k^2-m_c^2)^2}\right.\nonumber\\
&&\left. -\frac{g_s^2 (t^at^b)_{ij} G^a_{\alpha\beta}G^b_{\mu\nu}(f^{\alpha\beta\mu\nu}+f^{\alpha\mu\beta\nu}+f^{\alpha\mu\nu\beta}) }{4(k^2-m_c^2)^5}+\cdots\right\} \, ,\nonumber\\
f^{\lambda\alpha\beta}&=&(\!\not\!{k}+m_c)\gamma^\lambda(\!\not\!{k}+m_c)\gamma^\alpha(\!\not\!{k}+m_c)\gamma^\beta(\!\not\!{k}+m_c)\, ,\nonumber\\
f^{\alpha\beta\mu\nu}&=&(\!\not\!{k}+m_c)\gamma^\alpha(\!\not\!{k}+m_c)\gamma^\beta(\!\not\!{k}+m_c)\gamma^\mu(\!\not\!{k}+m_c)\gamma^\nu(\!\not\!{k}+m_c)\, ,
\end{eqnarray}
and  $t^n=\frac{\lambda^n}{2}$, the $\lambda^n$ is the Gell-Mann matrix \cite{Reinders85,Pascual-1984}.
In Eq.(6), we retain the term $\langle\bar{q}_j\sigma_{\mu\nu}q_i \rangle$  originate from the Fierz re-arrangement of the $\langle q_i \bar{q}_j\rangle$ to  absorb the gluons  emitted from other quark lines to  extract the mixed condensate $\langle\bar{q}g_s\sigma G q\rangle$ \cite{Wang-tetra-formula,WangHuangtao-2014-PRD}.

It is very difficult (or cumbersome) to carry out the integrals  both in the coordinate and momentum spaces directly due to appearance of the partial derives $\partial_\mu$ and  $\partial_\nu$. We perform integral by parts to exclude the terms proportional to the tensor structure $\frac{p_\mu p_\nu}{p^2}$, which only contributes to the scalar tetraquark states, and simplify the correlation function $\Pi_{\mu\nu}(p)$ greatly,   \begin{eqnarray}
\Pi_{\mu\nu}(p)&=&2i\varepsilon^{ijk}\varepsilon^{imn}\varepsilon^{i^{\prime}j^{\prime}k^{\prime}}\varepsilon^{i^{\prime}m^{\prime}n^{\prime}}\int d^4x e^{ip \cdot x}   \nonumber\\
&&\partial_\mu{\rm Tr}\left[ \gamma_5 C^{kk^{\prime}}(x)\gamma_5 CS^{jj^{\prime}T}(x)C\right] \partial_\nu{\rm Tr}\left[ \gamma_5 C^{n^{\prime}n}(-x)\gamma_5 C S^{m^{\prime}mT}(-x)C\right]\, .
\end{eqnarray}
Then we compute  the integrals both in the coordinate and momentum spaces,  and obtain the correlation function $\Pi(p^2)$ therefore the spectral density at the level of   quark-gluon degrees  of freedom.

 Once analytical expressions of the QCD spectral density  are obtained,  we can take the
quark-hadron duality below the continuum threshold  $s_0$ and perform Borel transform  with respect to
the variable $P^2=-p^2$ to obtain  the  QCD sum rules:
\begin{eqnarray}
\lambda^2_{Y}\, \exp\left(-\frac{M^2_{Y}}{T^2}\right)= \int_{4m_c^2}^{s_0} ds\, \rho(s) \, \exp\left(-\frac{s}{T^2}\right) \, ,
\end{eqnarray}
where
\begin{eqnarray}
\rho(s)&=&\rho_{0}(s)+\rho_{3}(s) +\rho_{4}(s)+\rho_{5}(s)+\rho_{6}(s)+\rho_{7}(s) +\rho_{8}(s)+\rho_{10}(s)\, ,
\end{eqnarray}

\begin{eqnarray}
\rho_{0}(s)&=&\frac{1}{61440\pi^6}\int dy dz\,\frac{yz\left(1-y-z\right)^4}{1-y} \left(s-\overline{m}_c^2\right)^4
\Big[s-\overline{m}_c^2 -2y\left(8s-3\overline{m}_c^2\right)\Big] \nonumber\\
&&-\frac{1}{12288\pi^6} \int dy dz\, \frac{yz^2\left(1-y-z\right)^3}{1-y}\left(s-\overline{m}_c^2\right)^4\left(3s-\overline{m}_c^2\right)\nonumber\\
&&+ \frac{1}{3840\pi^6}\int dy dz\, y^2z^2\left(1-y-z\right)^3 \left(s-\overline{m}_c^2\right)^3 \left(18s^2-16s\overline{m}_c^2+3\overline{m}_c^4\right) \nonumber\\
&&+\frac{1}{20480\pi^6} \int dy dz \, yz\left(1-y-z\right)^3 \left(s-\overline{m}_c^2\right)^4\Big[s-\overline{m}_c^2 +4y\left(8s-3\overline{m}_c^2\right)\Big]\, ,
\end{eqnarray}

\begin{eqnarray}
\rho_{3}(s)&=&-\frac{m_c\langle\bar{q}q\rangle}{48\pi^4}  \int dy dz\, yz\left(1-y-z\right) \left(s-\overline{m}_c^2\right)^2
\Big[s-\overline{m}_c^2+3y\left(2s-\overline{m}_c^2\right)\Big] \, ,
\end{eqnarray}

\begin{eqnarray}
\rho_{4}(s)&=&-\frac{m_c^2}{9216\pi^4}\langle\frac{\alpha_{s}GG}{\pi}\rangle \int dy dz\,\frac{z\left(1-y-z\right)^4}{y^2\left(1-y\right)}
\left(s-\overline{m}_c^2\right)\Big[s-\overline{m}_c^2 -2y\left(5s-3\overline{m}_c^2\right)\Big] \nonumber\\
&&+\frac{m_c^2}{9216\pi^4}\langle\frac{\alpha_{s}GG}{\pi}\rangle \int dy dz\, \frac{z^2\left(1-y-z\right)^3}{y^2\left(1-y\right)}
 \left(s-\overline{m}_c^2\right)\left(9s-5\overline{m}_c^2\right)  \nonumber\\
&&-\frac{m_c^2}{1152\pi^4}\langle\frac{\alpha_{s}GG}{\pi}\rangle \int dy dz\, \frac{z^2\left(1-y-z\right)^3}{y}
\left(15s^2-20s\overline{m}_c^2+6\overline{m}_c^4\right)\nonumber\\
&&-\frac{m_c^2}{1024\pi^4}\langle\frac{\alpha_{s}GG}{\pi}\rangle \int dy dz\, \frac{z\left(1-y-z\right)^3}{y^2} \left(s-\overline{m}_c^2\right)^2  \nonumber\\
&&+\frac{m_c^2}{3072\pi^4}\langle\frac{\alpha_{s}GG}{\pi}\rangle \int dy dz\, \left(\frac{z}{y^2}+\frac{y}{z^2}\right)\left(1-y-z\right)^3
\left(s-\overline{m}_c^2\right)\Big[s-\overline{m}_c^2 -2y\left(5s-3\overline{m}_c^2\right)\Big]\nonumber\\
&&+\frac{1}{6144\pi^4}\langle\frac{\alpha_{s}GG}{\pi}\rangle \int dy dz\, \frac{z\left(1-y-z\right)^3}{1-y}
\left(s-\overline{m}_c^2\right)^2\Big[s-\overline{m}_c^2 -6y\left(2s-\overline{m}_c^2\right)\Big] \nonumber\\
&&-\frac{1}{6144\pi^4}\langle\frac{\alpha_{s}GG}{\pi}\rangle \int dy dz\, \frac{z^2\left(1-y-z\right)^2}{1-y}\left(s-\overline{m}_c^2\right)^2\left(11s-5\overline{m}_c^2\right) \nonumber\\
&&+\frac{1}{256\pi^4}\langle\frac{\alpha_{s}GG}{\pi}\rangle  \int dy dz\, yz^2\left(1-y-z\right)^2\left(s-\overline{m}_c^2\right)
\left(7s^2-8s\overline{m}_c^2+2\overline{m}_c^4\right) \nonumber\\
&&-\frac{1}{2048\pi^4}\langle\frac{\alpha_{s}GG}{\pi}\rangle  \int dy dz\,  z\left(1-y-z\right)^2
\left(s-\overline{m}_c^2\right)^2\Big[s-\overline{m}_c^2 -6y\left(2s-\overline{m}_c^2\right)\Big] \, ,
\end{eqnarray}

\begin{eqnarray}
\rho_{5}(s)&=&\frac{m_c\langle\bar{q}g_{s}\sigma Gq\rangle}{64\pi^4} \int dy dz\, yz \left(s-\overline{m}_c^2\right)
\Big[s-\overline{m}_c^2+y\left(5s-3\overline{m}_c^2\right)\Big] \nonumber\\
&&+\frac{m_c\langle\bar{q}g_{s}\sigma Gq\rangle}{128\pi^4} \int dy dz\,y\left(1-y-z\right)\left(1-y\right)\left(s-\overline{m}_c^2\right)^2 \nonumber\\
&&-\frac{m_c\langle\bar{q}g_{s}\sigma Gq\rangle}{128\pi^4} \int dy dz\, y^2\left(1-y-z\right)\left(s-\overline{m}_c^2\right)\left(9s-5\overline{m}_c^2\right) \nonumber\\
&&-\frac{3m_c\langle\bar{q}g_{s}\sigma Gq\rangle}{128\pi^4} \int dy dz\, \left(y+z\right)\left(1-y-z\right) \left(s-\overline{m}_c^2\right)^2 \nonumber\\
&&+\frac{m_c \langle\bar{q}g_{s}\sigma Gq\rangle}{128\pi^4} \int dy dz\,
 y\left(1-y-z\right) \left(s-\overline{m}_c^2\right) \Big[s-\overline{m}_c^2-2y\left(5s-3\overline{m}_c^2\right)\Big]\ ,
\end{eqnarray}

\begin{eqnarray}
\rho_{6}(s)&=&\frac{m_c^2\langle\bar{q}q\rangle^2}{12\pi^2} \int dy\, y\left(1-y\right)\left(s-\widetilde{m}_c^2\right)\, ,
\end{eqnarray}

\begin{eqnarray}
\rho_{7}(s)&=&\frac{m_c^3\langle\bar{q}q\rangle}{144\pi^2} \langle\frac{\alpha_{s}GG}{\pi}\rangle  \int dy dz\,
\left(\frac{z}{y^2}+\frac{y}{z^2}\right)\left(1-y-z\right)\Big[1+3y +ys\,\delta\left(s-\overline{m}_c^2\right)\Big] \nonumber\\
&&-\frac{m_c\langle\bar{q}q\rangle}{48\pi^2} \langle\frac{\alpha_{s}GG}{\pi}\rangle  \int dy dz\,
\frac{y\left(1-y-z\right)}{z}\Big[s-\overline{m}_c^2+y\left(4s-3\overline{m}_c^2\right) \Big] \nonumber\\
&&+\frac{m_c \langle\bar{q}q\rangle}{192\pi^2} \langle\frac{\alpha_{s}GG}{\pi}\rangle  \int dy dz\, z
\Big[3\left(s-\overline{m}_c^2\right)-2y\left(4s-3\overline{m}_c^2\right)\Big]\nonumber\\
&& -\frac{m_c\langle\bar{q}q\rangle}{288\pi^2}\langle\frac{\alpha_{s}GG}{\pi}\rangle \int dy dz\, y\left(1-y\right)\left(s-\overline{m}_c^2\right) \nonumber\\
&&+\frac{m_c\langle\bar{q}q\rangle}{288\pi^2} \langle\frac{\alpha_{s}GG}{\pi}\rangle \int dy dz\, y^2 \left(7s-5\overline{m}_c^2\right) \nonumber\\
&&-\frac{m_c\langle\bar{q}q\rangle}{288\pi^2} \langle\frac{\alpha_{s}GG}{\pi}\rangle \int dy\, y\left(1-y\right)
\Big[ s-\widetilde{m}_c^2+y \left(4s-3\widetilde{m}_c^2\right)\Big]\, ,
\end{eqnarray}

\begin{eqnarray}
\rho_{8}(s)&=&-\frac{m_c^2\langle\bar{q}q\rangle  \langle\bar{q}g_{s}\sigma Gq\rangle}{24\pi^2} \int dy\, y\left(1-y\right)
\Big[3+s\,\delta\left(s-\widetilde{m}_c^2\right)\Big] +\frac{m_c^2\langle\bar{q}q\rangle  \langle\bar{q}g_{s}\sigma Gq\rangle}{24\pi^2} \int dy\, ,
\end{eqnarray}

\begin{eqnarray}
\rho_{10}(s)&=&\frac{203m_c^2 \langle\bar{q}g_{s}\sigma Gq\rangle^2}{9216\pi^2} \int dy\, \delta\left(s-\widetilde{m}_c^2\right)\nonumber\\
&&+\frac{m_c^2 \langle\bar{q}g_{s}\sigma Gq\rangle^2}{32\pi^2} \int dy\,  y\left(1-y\right) \left(1+\frac{2s}{3T^2}+\frac{s^2}{6T^4}\right)\delta\left(s-\widetilde{m}_c^2\right)\nonumber\\
&&-\frac{m_c^2\langle\bar{q}g_{s}\sigma Gq\rangle^2}{48\pi^2} \int dy\left(1+\frac{s}{2T^2}\right)\delta\left(s-\widetilde{m}_c^2\right) \nonumber\\
&& -\frac{m_c^4\langle\bar{q}q\rangle^2}{108T^2} \langle\frac{\alpha_{s}GG}{\pi}\rangle  \int dy\, \frac{1-y}{y^2}\delta\left(s-\widetilde{m}_c^2\right) \nonumber\\
&&+\frac{m_c^2\langle\bar{q}q\rangle^2}{36} \langle\frac{\alpha_{s}GG}{\pi}\rangle  \int dy\, \frac{1-y}{y}\delta\left(s-\widetilde{m}_c^2\right)\nonumber\\
&& -\frac{m_c^2\langle\bar{q}q\rangle^2}{108}\langle\frac{\alpha_{s}GG}{\pi}\rangle \int dy\,
\left(1+\frac{s}{2T^2}\right)\delta\left(s-\widetilde{m}_c^2\right) \nonumber\\
&&+ \frac{m_c^2\langle\bar{q}q\rangle^2}{36}\langle\frac{\alpha_{s}GG}{\pi}\rangle \int dy\,
 y\left(1-y\right) \left(1+\frac{2s}{3T^2}+\frac{s^2}{6T^4}\right)\delta\left(s-\widetilde{m}_c^2\right)\, ,
\end{eqnarray}
where $\int dydz=\int_{y_i}^{y_f}dy \int_{z_i}^{1-y}dz$, $y_{f}=\frac{1+\sqrt{1-4m_c^2/s}}{2}$,
$y_{i}=\frac{1-\sqrt{1-4m_c^2/s}}{2}$, $z_{i}=\frac{y
m_c^2}{y s -m_c^2}$, $\overline{m}_c^2=\frac{(y+z)m_c^2}{yz}$,
$ \widetilde{m}_c^2=\frac{m_c^2}{y(1-y)}$, $\int_{y_i}^{y_f}dy \to \int_{0}^{1}dy$, $\int_{z_i}^{1-y}dz \to \int_{0}^{1-y}dz$,  when the $\delta$ functions $\delta\left(s-\overline{m}_c^2\right)$ and $\delta\left(s-\widetilde{m}_c^2\right)$ appear.

 In this article, we carry out the operator product expansion up to the vacuum condensates of   dimension-10, and take into account the vacuum condensates which are
vacuum expectations  of the operators  of the orders $\mathcal{O}( \alpha_s^{k})$ with $k\leq 1$ consistently.
The condensates $\langle g_s^3 GGG\rangle$, $\langle \frac{\alpha_s GG}{\pi}\rangle^2$,
 $\langle \frac{\alpha_s GG}{\pi}\rangle\langle \bar{q} g_s \sigma Gq\rangle$ have the dimensions 6, 8, 9, respectively,  but they are   the vacuum expectations
of the operators of the order    $\mathcal{O}( \alpha_s^{3/2})$, $\mathcal{O}(\alpha_s^2)$, $\mathcal{O}( \alpha_s^{3/2})$, respectively, and are discarded \cite{Wang-tetra-formula,WangHuangtao-2014-PRD}.

We derive Eq.(9) with respect to  $\tau=\frac{1}{T^2}$, then eliminate the
 pole residue $\lambda_{Y}$, and obtain the QCD sum rules for
 the mass of the vector   tetraquark state $Y(4260/4220)$,
 \begin{eqnarray}
 M^2_{Y}&=& -\frac{\int_{4m_c^2}^{s_0} ds\frac{d}{d \tau}\rho(s)\exp\left(-\tau s \right)}{\int_{4m_c^2}^{s_0} ds \rho(s)\exp\left(-\tau s\right)}\, .
\end{eqnarray}

\section{Numerical results and discussions}
We take  the standard values of the vacuum condensates $\langle
\bar{q}q \rangle=-(0.24\pm 0.01\, \rm{GeV})^3$,   $\langle
\bar{q}g_s\sigma G q \rangle=m_0^2\langle \bar{q}q \rangle$,
$m_0^2=(0.8 \pm 0.1)\,\rm{GeV}^2$,  $\langle \frac{\alpha_s
GG}{\pi}\rangle=(0.33\,\rm{GeV})^4 $    at the energy scale  $\mu=1\, \rm{GeV}$
\cite{SVZ79,Reinders85,Colangelo-Review}, and choose the $\overline{MS}$ mass $m_{c}(m_c)=(1.275\pm0.025)\,\rm{GeV}$ from the Particle Data Group \cite{PDG}, and set $m_u=m_d=0$.
Moreover, we take into account the energy-scale dependence of  the input parameters on the QCD side,
\begin{eqnarray}
\langle\bar{q}q \rangle(\mu)&=&\langle\bar{q}q \rangle(Q)\left[\frac{\alpha_{s}(Q)}{\alpha_{s}(\mu)}\right]^{\frac{12}{25}}\, , \nonumber\\
 \langle\bar{q}g_s \sigma Gq \rangle(\mu)&=&\langle\bar{q}g_s \sigma Gq \rangle(Q)\left[\frac{\alpha_{s}(Q)}{\alpha_{s}(\mu)}\right]^{\frac{2}{25}}\, , \nonumber\\ m_c(\mu)&=&m_c(m_c)\left[\frac{\alpha_{s}(\mu)}{\alpha_{s}(m_c)}\right]^{\frac{12}{25}} \, ,\nonumber\\
\alpha_s(\mu)&=&\frac{1}{b_0t}\left[1-\frac{b_1}{b_0^2}\frac{\log t}{t} +\frac{b_1^2(\log^2{t}-\log{t}-1)+b_0b_2}{b_0^4t^2}\right]\, ,
\end{eqnarray}
   where $t=\log \frac{\mu^2}{\Lambda^2}$, $b_0=\frac{33-2n_f}{12\pi}$, $b_1=\frac{153-19n_f}{24\pi^2}$, $b_2=\frac{2857-\frac{5033}{9}n_f+\frac{325}{27}n_f^2}{128\pi^3}$,  $\Lambda=210\,\rm{MeV}$, $292\,\rm{MeV}$  and  $332\,\rm{MeV}$ for the flavors  $n_f=5$, $4$ and $3$, respectively  \cite{PDG,Narison-mix}, and evolve all the input parameters to the optimal energy scale   $\mu$ to extract the mass of the
   vector tetraquark state $Y(4260/4220)$.

 In this article, we search for the ideal  Borel parameter $T^2$ and continuum threshold parameter $s_0$  to satisfy   the  following four criteria:\\
$\bf 1.$ Pole dominance at the phenomenological side;\\
$\bf 2.$ Convergence of the operator product expansion;\\
$\bf 3.$ Appearance of the Borel platforms;\\
$\bf 4.$ Satisfying the energy scale formula,\\
 using try and error.

 In the four-quark system $q\bar{q}^{\prime}Q\bar{Q}$,
 the $Q$-quark serves as a static well potential and  combines with  the light quark $q$  to form a heavy diquark $\mathcal{D}$ in  color antitriplet or combines with the light antiquark $\bar{q}^\prime$  to form a heavy meson-like state or correlation (not a physical meson) in  color singlet,
while the $\bar{Q}$-quark serves  as another static well potential and combines with the light antiquark $\bar{q}^\prime$  to form a heavy antidiquark $\mathcal{\bar{D}}$ in  color triplet or combines with the light quark state $q$  to form another heavy meson-like state or correlation (not a physical meson) in  color singlet \cite{Wang-tetra-formula,Wang-tetra-NPA,WangHuang-molecule}.
 Then  the  $\mathcal{D}$ and $\mathcal{\bar{D}}$ combine with together   to form a compact tetraquark state, the two meson-like states (not two physical mesons)  combine together to form a  physical molecular state \cite{Wang-tetra-formula,Wang-tetra-NPA,WangHuang-molecule},
the two heavy quarks $Q$ and $\bar{Q}$ stabilize the tetraquark state \cite{Brodsky-PRL}. The tetraquark states $\mathcal{D\bar{D}}$  are characterized by the effective heavy quark masses ${\mathbb{M}}_Q$ and the virtuality $V=\sqrt{M^2_{X/Y/Z}-(2{\mathbb{M}}_Q)^2}$. It is natural to take the energy  scale $\mu=V=\sqrt{M^2_{X/Y/Z}-(2{\mathbb{M}}_Q)^2}$ \cite{Wang-tetra-formula,Wang-tetra-NPA,WangHuang-molecule}.
We cannot obtain energy scale independent QCD sum rules, but we have an energy scale formula to determine the energy scales consistently, which works well even for the  hidden-charm pentaquark states \cite{WangPentaQuark},  the updated value ${\mathbb{M}}_c=1.82\,\rm{GeV}$ \cite{WangEPJC-1601}.

 In Refs.\cite{Wang-tetra-formula,WangHuangtao-2014-PRD,Wang-tetra-NPA}, we study the hidden-charm or hidden-bottom tetraquark states, the heavy diquarks and heavy antidiquarks are in relative S-wave, if there exist  relative
 P-waves, the P-waves lie in between the heavy (anti)quark and light (anti)quark in the heavy (anti)diquark. In the present work, we study the vector tetraquark  state which has a relative P-wave between the charmed diquark and charmed antidiquark. If a relative P-wave costs  about $0.5\,\rm{GeV}$, then the energy scale formula is modified to be
\begin{eqnarray}
\mu&=&\sqrt{M^2_{Y}-(2{\mathbb{M}}_c+0.5\,\rm{GeV})^2}=\sqrt{M^2_{Y}-(4.1\,\rm{GeV})^2}\, .
\end{eqnarray}
In calculations, we observe that if we take the continuum threshold parameter $\sqrt{s_0}=4.8\pm 0.1\,\rm{GeV}$, Borel parameter $T^2=(2.2-2.8)\,\rm{GeV}^2$, energy scale $\mu=1.1\,\rm{GeV}$, the pole contribution of the ground state vector tetraquark state $Y(4260/4220)$ is about $(49-81)\%$, the predicted mass is about $M_{Y}=4.24\,\rm{GeV}$, the modified energy scale formula is well satisfied.

In Fig.1, we plot the pole contribution with variation of the Borel parameter, from the figure, we can see that the pole contribution decreases  monotonously with increase of the Borel parameter, the pole contribution reaches about $50\%$ at the point  $T^2=2.8\,\rm{GeV^2}$ and $\sqrt{s_0}=4.7\,\rm{GeV}$, we can obtain the upper  bound $T^2_{max}=2.8\,\rm{GeV^2}$.  In Fig.2, we plot the  contributions of the vacuum condensates of dimension $n$ in the operator product expansion, which are defined by
\begin{eqnarray}
D(n)&=& \frac{  \int_{4m_c^2}^{s_0} ds\,\rho_{n}(s)\,\exp\left(-\frac{s}{T^2}\right)}{\int_{4m_c^2}^{s_0} ds \,\rho(s)\,\exp\left(-\frac{s}{T^2}\right)}\, .
\end{eqnarray}
From the figure, we can see that the contributions of the vacuum condensates of dimensions $3$, $5$, $6$ and $8$ are very large, and change quickly with variation of the Borel parameter $T^2$ at the region $T^2<2.2\,\rm{GeV}^2$,    the operator product expansion is not convergent, we can obtain the lower bound   $T^2_{min}=2.2\,\rm{GeV^2}$. At the region $T^2\geq2.2\,\rm{GeV^2}$, the contribution of the vacuum condensate of dimension $n=3$ is large, but the contributions of the vacuum condensates of dimensions $3, \,5,\,6,\,8$ have the hierarchy $D(3)\gg |D(5)|\sim D(6)\gg |D(8)|$, the contributions of the vacuum condensates of the dimensions $4$, $7$, $10$ are tiny,  the operator product expansion is  convergent.
 The Borel window is $T^2=(2.2-2.8)\,\rm{GeV}^2$, where operator product expansion is well convergent.

We take into account all uncertainties of the input parameters,
and obtain the values of the mass and pole residue of
 the   vector tetraquark state $Y(4260/4220)$, which are  shown explicitly in Figs.3-4,
\begin{eqnarray}
M_{Y}&=&4.24\pm0.10\,\rm{GeV} \, ,  \nonumber\\
\lambda_{Y}&=&\left( 2.31 \pm0.45 \right) \times 10^{-2}\,\rm{GeV}^6 \,   .
\end{eqnarray}
From Figs.3-4, we can see that  there appear platforms in the Borel window. Now the four criteria of the QCD sum rules are all satisfied, and we expect to make reliable predictions.

 The predicted mass $M_{Y}=4.24\pm0.10\,\rm{GeV}$ is in excellent agreement with the experimental value  $M_{Y(4220)}=4222.0\pm3.1\pm 1.4\,  \rm{MeV}$ from the BESIII    collaboration \cite{BES-Y4220-Y4320}, or the experimental value  $M_{Y(4260)}=4230.0\pm 8.0\,  \rm{MeV}$ from Particle Data Group \cite{PDG}, which supports  assigning the $Y(4260/4220)$  to be the  $C\gamma_5\otimes\stackrel{\leftrightarrow}{\partial}_\mu\otimes \gamma_5C$   type vector tetraquark state. The average value of the width of the $Y(4260)$ is $55\pm19\,\rm{MeV}$, the relative P-wave between the diquark and antidiquark disfavors rearrangement of the quarks to form meson pairs, which
 can account for the small width.

From Fig.3, we can see that the mass $M_{Z_c(4100)}$ lies below the lower bound of the predicted mass of the $C\gamma_5\otimes\stackrel{\leftrightarrow}{\partial}_\mu\otimes \gamma_5C$ type vector tetraquark state  $c\bar{c}q\bar{q}$, which disfavors assigning the $Z_c(4100)$  to be the  $C\gamma_5\otimes\stackrel{\leftrightarrow}{\partial}_\mu\otimes \gamma_5C$  type vector tetraquark state  $c\bar{c}q\bar{q}$.

In Refs.\cite{Wang-3915-CgmCgm,Wang-3915-C5C5}, we study the $C\gamma_\mu\otimes \gamma^\mu C$-type, $C\gamma_\mu\gamma_5\otimes \gamma_5\gamma^\mu C$-type,
$C\gamma_5\otimes \gamma_5 C$-type, $C\otimes  C$-type $cs\bar{c}\bar{s}$ scalar tetraquark states with the QCD sum rules in a systematic way, and obtain the predictions $M_{C\gamma_\mu\otimes \gamma^\mu C}=3.92^{+0.19}_{-0.18}\,\rm{GeV}$ and $M_{C\gamma_5\otimes \gamma_5C}=3.89\pm0.05\,\rm{GeV}$, which support assigning the $X(3915)$ to be the $C\gamma_\mu\otimes \gamma^\mu C$-type or $C\gamma_5\otimes \gamma_5 C$-type $cs\bar{c}\bar{s}$ scalar tetraquark state.
In fact, the $SU(3)$ breaking effects of the masses of the $cs\bar{c}\bar{s}$ and $cq\bar{c}\bar{q}$  tetraquark states from the QCD sum rules are rather  small, if the scalar tetraquark state $cq\bar{c}\bar{q}$ has the mass $M_{C\gamma_\mu\otimes \gamma^\mu C}=3.92^{+0.19}_{-0.18}\,\rm{GeV}$, which    is compatible with the LHCb data  $M_{Z_c}=4096 \pm 20^{+18}_{-22}\,\rm{MeV}$ and $\Gamma_{Z_c}= 152 \pm 58^{+60}_{-35}\,\rm{MeV}$ considering the uncertainties \cite{LHCb-Z4100}, and favors assigning the $Z_c(4100)$ to be  the $C\gamma_\mu\otimes \gamma^\mu C$-type scalar tetraquark state.

In Ref.\cite{WangY4360Y4660-1803}, we choose the $C\otimes \gamma_\mu C$ type and
$C\gamma_5 \otimes \gamma_5\gamma_\mu C$ type vector currents to study the vector tetraquark states, the net effects of the relative P-waves   are embodied in the underlined  $\gamma_5$ in the  $C\gamma_5 \underline{\gamma_5} \otimes \gamma_\mu C$ type and
$C\gamma_5 \otimes \underline{\gamma_5}\gamma_\mu C$ type currents or in the underlined  $\gamma^\alpha$ in the  $C\gamma_\alpha \underline{\gamma^\alpha} \otimes \gamma_\mu C$ type  currents, and obtain the masses  $M_{C\otimes \gamma_\mu C}=4.59\pm0.08\,\rm{GeV}$ and $M_{C\gamma_5 \otimes \gamma_5\gamma_\mu C}=4.34\pm0.08\,\rm{GeV}$. The $C \otimes \gamma_\mu C$ type tetraquark states have larger masses than the corresponding $C\gamma_5 \otimes \gamma_5\gamma_\mu C$  type tetraquark states, as $C \otimes \gamma_\mu C=\left[C\gamma_5 \underline{\gamma_5} \otimes \gamma_\mu C\right]\oplus \left[C\gamma_\alpha \underline{\gamma^\alpha} \otimes \gamma_\mu C\right]$  and $C\gamma_5 \otimes \gamma_5\gamma_\mu C=C\gamma_5 \otimes \underline{\gamma_5}\gamma_\mu C$, the $C\gamma_\mu$ diquark states have slightly  larger masses than the corresponding $C\gamma_5$ diquark states from the QCD sum rules \cite{WangDiquark,WangLDiquark}.
The vector tetraquark  masses $M_{C\otimes \gamma_\mu C}$ and $M_{C\gamma_5 \otimes \gamma_5\gamma_\mu C}$ differ from the vector tetraquark mass $M_{C\gamma_5\otimes\stackrel{\leftrightarrow}{\partial}_\mu\otimes \gamma_5C}$ greatly. For the conventional ground state $c\bar{q}$ mesons, the energy gaps between the S-wave and P-wave states are about $0.5\,\rm{GeV}$, if the relative P-waves between the $q$-quark and $c$-quark in the diquark states $cq$ cost about $0.5\,\rm{GeV}$ \cite{PDG},   
the masses of the  $\stackrel{\leftrightarrow}{\partial}_\mu C\gamma_5\otimes \gamma_5C$ type, $C\gamma_5\otimes\stackrel{\leftrightarrow}{\partial}_\mu \gamma_5C$ type,  $\stackrel{\leftrightarrow}{\partial}_\mu C\gamma_\alpha \otimes \gamma^\alpha C$ type and $C\gamma_\alpha \otimes \stackrel{\leftrightarrow}{\partial}_\mu \gamma^\alpha C$ vector tetraquark states are estimated to be $4.4\,\rm{GeV}$ according the $C\gamma_\mu\otimes \gamma^\mu C$-type and  
$C\gamma_5\otimes \gamma_5 C$-type scalar tetraquark  masses \cite{Wang-3915-CgmCgm,Wang-3915-C5C5}, which differs from the present prediction $M_{C\gamma_5\otimes\stackrel{\leftrightarrow}{\partial}_\mu\otimes \gamma_5C}=4.24\pm0.10\,\rm{GeV}$ greatly. Before draw a definite conclusion, we should study the
masses of the  $\stackrel{\leftrightarrow}{\partial}_\mu C\gamma_5\otimes \gamma_5C$ type, $C\gamma_5\otimes\stackrel{\leftrightarrow}{\partial}_\mu \gamma_5C$ type,  $\stackrel{\leftrightarrow}{\partial}_\mu C\gamma_\alpha \otimes \gamma^\alpha C$ type and $C\gamma_\alpha \otimes \stackrel{\leftrightarrow}{\partial}_\mu \gamma^\alpha C$ vector tetraquark states with the QCD sum rules directly, this is our next work.

\begin{figure}
 \centering
 \includegraphics[totalheight=6cm,width=9cm]{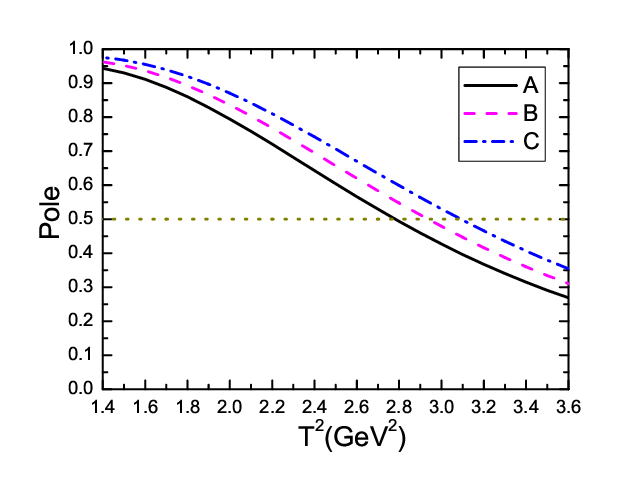}
 \caption{ The pole contributions  with variation of the Borel parameter $T^2$,  where the $A$, $B$ and $C$ denote the threshold parameters $\sqrt{s_0}=4.7\,\rm{GeV}$, $4.8\,\rm{GeV}$ and $4.9\,\rm{GeV}$, respectively.  }
\end{figure}

\begin{figure}
 \centering
 \includegraphics[totalheight=6cm,width=9cm]{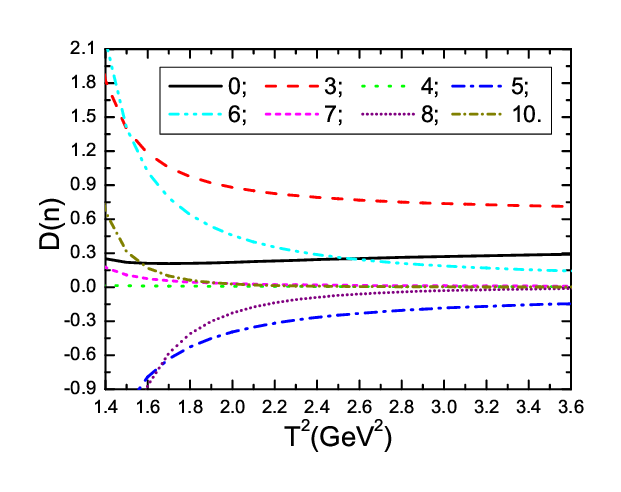}
 \caption{ The contributions of the vacuum condensates of dimension $n$ with variation of the Borel parameter $T^2$ for the threshold parameter $\sqrt{s_0}=4.8\,\rm{GeV}$.  }
\end{figure}

\begin{figure}
 \centering
 \includegraphics[totalheight=6cm,width=9cm]{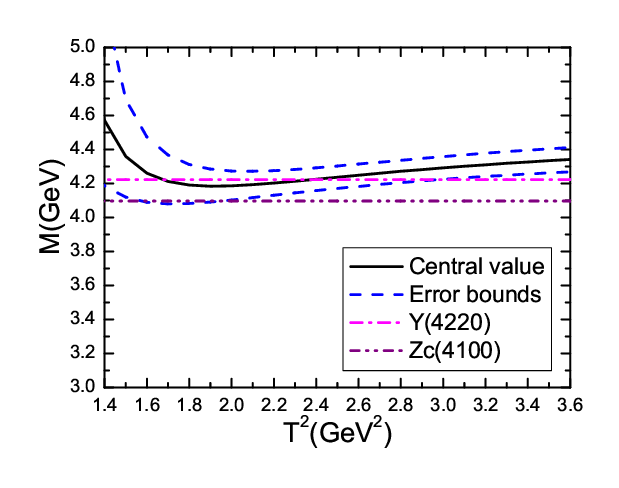}
 \caption{ The mass  of the $Y(4260/4220)$  as vector tetraquark state with variation of the Borel parameter $T^2$.  }
\end{figure}

\begin{figure}
 \centering
 \includegraphics[totalheight=6cm,width=9cm]{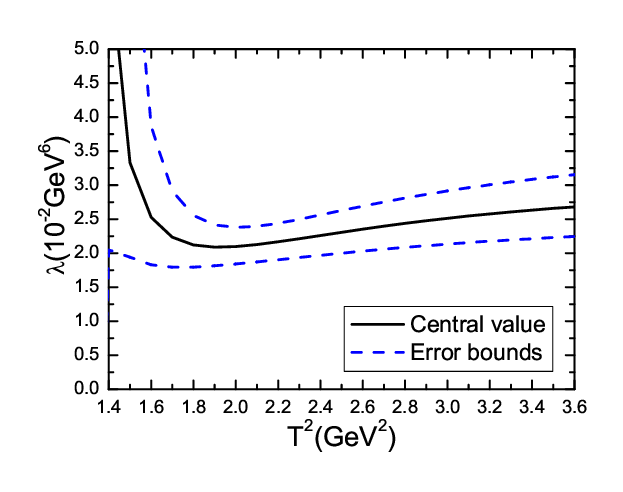}
 \caption{ The pole residue  of the $Y(4260/4220)$ as vector tetraquark state  with variation of the Borel parameter $T^2$.  }
\end{figure}

\section{Conclusion}
In this article, we take the $Y(4260/4220)$ as the vector tetraquark state with $J^{PC}=1^{--}$, and construct  the $C\gamma_5\otimes\stackrel{\leftrightarrow}{\partial}_\mu\otimes \gamma_5C$ type current to study its mass and pole residue with the QCD sum rules in details  by taking into account the vacuum condensates up to dimension 10 in a consistent way in the operator product expansion, and use the modified  energy scale formula $\mu=\sqrt{M^2_{X/Y/Z}-(2{\mathbb{M}}_c+0.5\rm{GeV})^2}$ with the effective $c$-quark mass ${\mathbb{M}}_c$ to determine the optimal energy scale of the QCD spectral density. The  predicted mass $M_{Y}=4.24\pm0.10\,\rm{GeV}$ is in excellent agreement with the experimental value  $M_{Y(4220)}=4222.0\pm3.1\pm 1.4\,  \rm{MeV}$ from the BESIII    collaboration or the experimental value  $M_{Y(4260)}=4230.0\pm 8.0\,  \rm{MeV}$ from Particle Data Group, and supports assigning the $Y(4260/4220)$  to be the  $C\gamma_5\otimes\stackrel{\leftrightarrow}{\partial}_\mu\otimes \gamma_5C$  type vector tetraquark state, and disfavors assigning the $Z_c(4100)$  to be the  $C\gamma_5\otimes\stackrel{\leftrightarrow}{\partial}_\mu\otimes \gamma_5C$  type vector tetraquark state.  It is the first time that the QCD sum rules have reproduced  the mass of the $Y(4260/4220)$ as a vector tetraquark state.

\section*{Acknowledgements}
This  work is supported by National Natural Science Foundation, Grant Number  11775079.

\end{document}